\documentstyle[12pt,epsfig]{article}  
\begin{document} 
\baselineskip=20pt

\def\la{\mathrel{\mathpalette\fun <}}
\def\ga{\mathrel{\mathpalette\fun >}}
\def\fun#1#2{\lower3.6pt\vbox{\baselineskip0pt\lineskip.9pt
\ialign{$\mathsurround=0pt#1\hfil##\hfil$\crcr#2\crcr\sim\crcr}}} 

\begin{titlepage} 
\begin{center}
{\Large \bf Double parton correlations versus factorized distributions  } \\

\vspace{4mm}
V.L.~Korotkikh, A.M.~Snigirev  \\
M.V.Lomonosov Moscow State University, D.V.Skobeltsyn Institute of Nuclear 
Physics \\
119992, Vorobievy Gory, Moscow, Russia \\ E-mail:~~vlk@lav01.sinp.msu.ru,
~~snigirev@lav01.sinp.msu.ru 
\end{center}  

\begin{abstract} 
Using the
generalized Lipatov-Altarelli-Parisi-Dokshitzer
equations for the two-parton distribution functions we show numerically that the
dynamical correlations contribute to these functions quite a lot in comparison
with the factorized components. At the scale of CDF hard process ($\sim 5$ GeV)
this contribution to the double gluon-gluon distribution 
 is nearly 10$\%$  and increases right up to 30$\%$ at the LHC 
scale ($\sim 100$ GeV) for the longitudinal momentum fractions $x \leq 0.1$
accessible to these measurements. For the finite longitudinal momentum
fractions $x \sim 0.2 \div 0.4$ the correlations are large right up to 90$\%$ in
accordance with the predicted QCD asymptotic behaviour.

\end{abstract}

\bigskip

\noindent 
$PACS$: ~~12.38.--t \\ 
$Keywords$: many parton distributions, leading logarithm  approximation

\end{titlepage}   
\newpage 

Recent CDF measurements~\cite{cdf} of the inclusive cross section for  double 
parton scattering have provided  new  and complementary information on the 
structure of the proton and possible parton-parton
correlations. Both the absolute rate for the double parton process and any
dynamics that correlations may introduce are therefore of interest. 
The possibility of observing two separate hard  collisions has been
proposed since long~\cite{landshoff,goebel}, and from that has also 
developed in a number of works~\cite{takagi,paver, humpert, odorico, 
sjostrand, trelani, trelani2,del}. The Tevatron and specially LHC allow us to obtain
huge data samples of these multiple interactions and to answer to many
challenging questions of yet poorly-understood aspects of QCD. A brief review of
the current situation and some progress in the modeling account of 
correlated flavour, colour,
longitudinal and transverse momentum distributions can be found
in ref.~\cite{sjostrand2}. Multiple interactions require an ansatz for the structure
of the incoming beams, i.e. correlations between the constituent partons. As a
simple ansatz, usually,  the two-parton distributions are supposed to be 
the product of two single-parton distributions times a momentum conserving 
phase space factor. In recent paper~\cite{snig03} it has been shown that this
hypothesis is in some contradiction with the leading logarithm approximation
of perturbative QCD  (in the framework of
which a parton model, as a matter of fact, was established in the quantum field 
theories~\cite{gribov,lipatov,dok}). Namely, 
the two-parton distribution functions being the
product of  two single distributions at some reference scale become to be
dynamically correlated at any different scale of a hard process. The value of
these correlations in comparison with the factorized components is the main
purpose of this Letter.

In order to be clear and to introduce the denotations
let us recall that,
for instance, the
differential cross section for the four-jet process (due to the simultaneous
interaction of two parton pairs) is given by~\cite{humpert, odorico}  
\begin{equation} 
\label{fourjet}
d \sigma = \sum \limits_{q/g} \frac{ d \sigma_{12} ~d \sigma_{34}}
{\sigma_{eff}}~ D_ p(x_1,x_3)~D_{\bar{p}}(x_2,x_4), 
\end{equation} 
where $d \sigma_{ij}$ stands for the two-jet cross section. The dimensional
factor $\sigma_{eff}$
in the denominator represents the total inelastic cross section which is an
estimate of the size of the hadron, $\sigma_{eff} ~\simeq~2 \pi r_ p^2$ (the factor
2 is introduced due to the identity of the two parton processes). With
the effective cross section measured by CDF, $(\sigma_{eff})_{CDF}=
(14.5 \pm 1.7^{+1.7}_{-2.3})$ mb~\cite{cdf}, one can estimate the transverse
size $r_p~\simeq 0.5$ fm, which is too small in comparison with the proton
radius $R_p$
extracted from $ep$ elastic scattering experiments. The relatively small
value of $(\sigma_{eff})_{CDF}$ with respect to the naive expectation
$2 \pi R_ p^2$ was, in fact, considered~\cite{trelani,trelani2} as  evidence of
nontrivial correlation effects in transverse space. But, apart from these
correlations, the longitudinal momentum correlations can also exist and 
they were   investigated in ref.~\cite{snig03}. The factorization ansatz is just applied
to the two-parton distributions incoming in eq.~(\ref{fourjet}):
\begin{equation} 
\label{factoriz}
D_ p(x_i,x_j)~ = ~ D_ p(x_i,Q^2)~ D_ p(x_j,Q^2)~(1-x_i-x_j),
\end{equation} 
where  $D_ p(x_i,Q^2)$ are the single quark/gluon momentum distributions at
the scale $Q^2$ (deter\-mi\-ned by a hard process).

However many parton distribution functions satisfy the generalized 
Lipatov-Altarelli-Parisi-Dokshitzer evolution equations derived for the first
time in refs~\cite{kirschner,snig} as well as single
parton distributions satisfy more known and cited Altarelli-Parisi
equations~\cite{lipatov,dok,altarelli}.
Under certain initial conditions these generalized
equations lead to  solutions, which are identical with the jet calculus rules
proposed originally for multiparton fragmentation functions by
Konishi-Ukawa-Veneziano~\cite{konishi} and are in some contradiction with the
factorization hypothesis (\ref{factoriz}). Here one should note that at the parton  
level this  is the strict assertion within the leading logarithm approximation. 

After introducing the natural dimensionless variable
$$t = \frac{1}{2\pi b} \ln \Bigg[1 + \frac{g^2(\mu^2)}{4\pi}b
\ln\Bigg(\frac{Q^2}{\mu^2}\Bigg)\Bigg]~=~\frac{1}{2\pi b}\ln\Bigg
[\frac{\ln(\frac{Q^2}{\Lambda^2_{QCD} })}
{\ln(\frac{\mu^2}{\Lambda^2_{QCD}})}\Bigg]
,~~~~~b = \frac{33-2n_f}{12\pi}~~
{\rm {in~ QCD}},$$
where $g(\mu^2)$ is the running coupling constant at the reference scale
$\mu^2$, $n_f$ is the number of active flavours, 
$\Lambda_{QCD}$ is the dimensional QCD 
parameter,
the Altarelli-Parisi equations read~\cite{lipatov,dok,altarelli}
\begin{equation}
\label{e1singl}
 \frac{dD_i^j(x,t)}{dt} = 
\sum\limits_{j{'}} \int \limits_x^1
\frac{dx{'}}{x{'}}D_i^{j{'}}(x{'},t)P_{j{'}\to j}\Bigg(\frac{x}{x{'}}\Bigg).
\end{equation}
\noindent
They describe the scaling violation 
of the parton distributions $ D^j_i(x,t)$ inside a dressed quark or gluon ($i,
j~=~q/g$).

We will not write  the kernels $P$ explicitly and  derive the generalized 
equations for  two-parton distributions
$D_i^{j_1j_2}(x_1,x_2,t)$, representing the probability that in a dressed
constituent $i$ one finds two bare partons  of  types $j_1$ and $j_2$ with 
the given longitudinal momentum fractions $x_1$ and $x_2$  (referring to
 ~\cite{snig03,lipatov, dok,kirschner, snig, altarelli} for details), we give
only their solutions via the convolution of single 
distributions~\cite{kirschner, snig}
\begin{eqnarray}
\label{solution}
& D_i^{j_1j_2}(x_1,x_2,t) = \\
& \sum\limits_{j{'}j_1{'}j_2{'}} \int\limits_{0}^{t}dt{'}
\int\limits_{x_1}^{1-x_2}\frac{dz_1}{z_1}
\int\limits_{x_2}^{1-z_1}\frac{dz_2}{z_2}~
D_i^{j{'}}(z_1+z_2,t{'}) \frac{1}{z_1+z_2}P_{j{'} \to
j_1{'}j_2{'}}\Bigg(\frac{z_1}{z_1+z_2}\Bigg) D_{j_1{'}}^{j_1}(\frac{x_1}{z_1},t-t{'}) 
D_{j_2{'}}^{j_2}(\frac{x_2}{z_2},t-t{'}).\nonumber
\end{eqnarray}
\noindent 
This convolution coincides with the jet calculus rules~\cite{konishi} as
mentioned above and is the  generalization of 
the  well-known Gribov-Lipatov relation
installed for single functions~\cite{gribov, dok} (the distribution
of bare partons inside a dressed constituent  is identical to the distribution 
of dressed constituents in the fragmentation of 
a bare parton in the leading logarithm 
approximation). 
The solution (\ref{solution}) shows that the distribution of partons is 
{\it {correlated}} in the leading logarithm approximation:
\begin{eqnarray}
\label{nonfact}
D_i^{j_1j_2}(x_1,x_2,t) \neq D_{i}^{j_1}(x_1,t) 
D_{i}^{j_2}(x_2,t).
\end{eqnarray}

Of course, it is interesting to find out the phenomenological issue of this
parton level consideration. This can be done within the well-known
factorization of soft and hard stages (physics of short and long
distances)~\cite{collins}. As a result the equations (\ref{e1singl}) 
 describe the evolution of parton distributions in a hadron with
$t ~(Q^2)$, if one replaces the index $i$ by index $h$ only. However, the initial
conditions for new equations at $t=0 ~(Q^2=\mu^2)$ are unknown a priori and must
be introduced phenomenologically or must be extracted from experiments 
or some models dealing with physics of long distances [at the parton level: 
$D_{i}^{j}(x,t=0)~= ~\delta_{ij} \delta(x-1)$; ~$D_i^{j_1j_2}(x_1,x_2,t=0)~=~0$].
Nevertheless the solution of the generalized  
Lipatov-Altarelli-Parisi-Dokshitzer evolution equations with the given initial
condition may be written as before via the convolution of single
distributions~\cite{snig03,snig}
\begin{eqnarray}
\label{solution1}
& D_h^{j_1j_2}(x_1,x_2,t)~ = ~ D_{h(QCD)}^{j_1j_2}(x_1,x_2,t)~+\\
& \sum\limits_{j_1{'}j_2{'}} 
\int\limits_{x_1}^{1-x_2}\frac{dz_1}{z_1}
\int\limits_{x_2}^{1-z_1}\frac{dz_2}{z_2}~
D_h^{j_1{'}j_2{'}}(z_1,z_2,0) D_{j_1{'}}^{j_1}(\frac{x_1}{z_1},t) 
D_{j_2{'}}^{j_2}(\frac{x_2}{z_2},t) ~,\nonumber
\end{eqnarray}
where
\begin{eqnarray}
\label{solQCD}
& D_{h(QCD)}^{j_1j_2}(x_1,x_2,t) =\\
& \sum\limits_{j{'}j_1{'}j_2{'}} \int\limits_{0}^{t}dt{'}
\int\limits_{x_1}^{1-x_2}\frac{dz_1}{z_1}
\int\limits_{x_2}^{1-z_1}\frac{dz_2}{z_2}~
D_h^{j{'}}(z_1+z_2,t{'}) \frac{1}{z_1+z_2}P_{j{'} \to
j_1{'}j_2{'}}\Bigg(\frac{z_1}{z_1+z_2}\Bigg) D_{j_1{'}}^{j_1}(\frac{x_1}{z_1},t-t{'}) 
D_{j_2{'}}^{j_2}(\frac{x_2}{z_2},t-t{'})\nonumber
\end{eqnarray}
are the dynamically correlated distributions given by perturbative QCD (compare 
(\ref{solution}) with (\ref{solQCD})).

The reckoning for the unsolved confinement problem (physics of long distances) is
the unknown nonperturbative two-parton correlation function $ D_h^{j_1{'}j_2{'}}(z_1,z_2,0)$
at some scale $\mu^2$. One can suppose that this function is the product of two
single-parton distributions times a momentum conserving 
factor at this scale $\mu^2$:
\begin{eqnarray}
\label{fact}
D_h^{j_1j_2}(z_1,z_2,0) ~=~ D_{h}^{j_1}(z_1,0) 
D_{h}^{j_2}(z_2,0)\theta(1-z_1-z_2).
\end{eqnarray}
\noindent
Then
\begin{eqnarray}
\label{solution2}
& D_h^{j_1j_2}(x_1,x_2,t)~ =~ D_{h(QCD)}^{j_1j_2}(x_1,x_2,t)~+~ 
\theta (1-x_1-x_2)\bigg(D_{h}^{j_1}(x_1,t)D_{h}^{j_2}(x_2,t)~+~\\
& \sum\limits_{j_1{'}j_2{'}} 
\int\limits_{x_1}^{1}\frac{dz_1}{z_1}
\int\limits_{x_2}^{1}\frac{dz_2}{z_2}~
D_h^{j_1{'}}(z_1,0)D_h^{j_2{'}}(z_2,0)
 D_{j_1{'}}^{j_1}(\frac{x_1}{z_1},t) 
D_{j_2{'}}^{j_2}(\frac{x_2}{z_2},t)[\theta(1-z_1-z_2)-1]
\bigg),
\nonumber \\
\end{eqnarray}
where
\begin{equation}
\label{1solution}
 D_h^j(x,t) = 
\sum\limits_{j{'}} \int \limits_x^1
\frac{dz}{z}~D_h^{j{'}}(z,0)~D_{j{'}}^j(\frac{x}{z},t)
\end{equation}
\noindent
is the solution of eq.~(\ref{e1singl}) with the given initial condition
$D_h^j(x,0)$ for parton distributions inside a hadron expressed via
distributions at the parton level. 

 This result~(\ref{solution2}) shows that
if the two-parton distributions are factorized at some scale $\mu^2$, then
the evolution violates this factorization {\it{ inevitably}} at any different
scale ($Q^2 \neq \mu^2$), apart from the violation due to 
the kinematic correlations induced by the momentum
conservation (given by $\theta$ functions){\footnote{  This is the analogue of
the  momentum conserving phase space factor in eq.~(\ref{factoriz})}}.

For a practical employment it is interesting to know the degree of this
violation. 
Partly this problem was investigated theoretically in refs.~\cite{snig, snig2}
and  for the two-particle correlations of fragmentation
functions in ref.~\cite{puhala}. That technique is based on the Mellin 
transformation of distribution functions as
\begin{eqnarray}
\label{mellin}
M_h^{j}(n,t) ~=~ \int\limits_{0}^{1}dx ~x^n~D_{h}^{j}(x,t). 
\end{eqnarray}
\noindent
After that the integrodifferential equations (\ref{e1singl}) 
 become systems of ordinary linear-differential equations of
first order with constant coefficients and can be solved explicitly~\cite{snig,
snig2}. In order to obtain the distributions in $x$ representation an inverse
Mellin transformation must be performed 
\begin{eqnarray}
\label{mellin in}
D_h^{j}(x,t) ~=~ \int\frac {dn}{2\pi i} ~x^{-n}~M_{h}^{j}(n,t), 
\end{eqnarray}
\noindent
where the integration runs along the imaginary axis to the right from all
$n$ singularities. This can be done numerically. However the asymptotic
behaviour can be estimated. Namely, with the growth of $t~(Q^2)$ the first term in
eq.~(\ref{solution1}) becomes {\it {dominant}} 
{\footnote{Such domination is the mathematical consequence of the relation between 
the maximum eigenvalues $\lambda(n)$ in the moments representation (after Mellin
transformation): ~$\lambda(n_1+n_2)~>~\lambda(n_1)+\lambda(n_2)$ in QCD at
the large $n_1,n_2$ (finite $x_1,x_2$), because  $\lambda(n)\sim -\ln(n),
n \gg 1$.}}
for  finite $x_1$ and $x_2$~\cite{snig2}.  
Thus the two-parton distribution functions 
``forget'' the initial conditions unknown a
priori and the correlations perturbatively calculated appear.

The asymptotic prediction ``teaches'' us a tendency only and tells nothing about the 
values of $x_1,x_2, t(Q^2)$ beginning from which the correlations are significant 
(the more so since the asymptotic behaviour takes place over the double logarithm
dimensionless variable $t$ as a function of $Q^2$). Naturally numerical estimations
can give an answer to this specific question. We do it using the CTEQ 
fit~\cite{cteq} for single distributions as an input in eq.~(\ref{solQCD}).
The nonperturbative initial conditions $D_h^j(x,0)$ are specified in a parametrized
form at a fixed low-energy scale $Q_0=\mu=1.3$ GeV. The particular function forms and
the value of $Q_0$ are not crucial for the CTEQ global analysis at the flexible
enough parametrization, which reads~\cite{pumplin} 
\begin{eqnarray}
\label{paramet}
x D_p^{j}(x,0) ~=~ A_0^j x^{A_1^j} (1 - x)^{A_2^j} e^{A_3^j x} (1 + e^{A_4^j}
x)^{A_5^j}.
\end{eqnarray}
The independent parameters $A_0^j,~ A_1^j,~ A_2^j,~ A_3^j, ~A_4^j, ~A_5^j$ for parton
flavour combinations $u_v \equiv u-\bar{u}$, $d_v \equiv d-\bar{d}$, $g$ and
$\bar{u}+\bar{d}$ are given in Appendix A of ref.~\cite{pumplin}. To distinguish the
$\bar{u}$ and $\bar{d}$ distributions the ratio $\bar{d} / \bar{u}$ is parametrized
as a sum of two terms:
\begin{eqnarray}
\label{paramet2}
 D_p^{\bar{d}}(x,0) \big/ D_p^{\bar{u}}(x,0)~=
 ~ A_0 x^{A_1} (1 - x)^{A_2} ~+~ (1 + A_3 x) (1 - x)^{A_4}
\end{eqnarray}
with the coefficients 
$A_0,~ A_1,~ A_2,~ A_3, ~A_4$ again from ref.~\cite{pumplin}.
The initial conditions for strange quarks are assumed:
$$D_p^{\bar{s}}(x,0) ~=~ D_p^{s}(x,0)~=~
0.2\Big(D_p^{\bar{u}}(x,0)~+~D_p^{\bar{d}}(x,0)\Big).$$

The parton distribution functions $D_p^{j}(x,t)$ at all higher $Q(t)$ are determined
from the input initial conditions $D_p^{j}(x,0)$ by the Altarelli-Parisi evolution
equations. The CTEQ Evolution package~\cite{Evolve} was used and adapted by us in
order to obtain numerically single distributions $D_i^{j}(x,t)$ at all $t$ and
at the parton level also. We fixed the fundamental parameter of perturbative QCD,
$\Lambda_{QCD} = 0.281$ GeV, that is in accordance with the strong coupling constant,
$\alpha_s(M_Z)~\simeq~0.2$, at the $Z$ resonance in one-loop approximation. 
Only the light quarks $u, d, s$ ($n_f = 3$) are taken into account in the
evolution equations and are treated as massless,
as usual. After that the triple integral (\ref{solQCD}) was calculated numerically for 
three values of $Q = 5, 100, 250$ GeV as a function of $x = x_1 = x_2$. To be specific
we considered the double gluon-gluon distribution function in the proton. In this case
only the kernel $P_{g \to g g}$ can be taken into account as giving the main
contribution to the perturbative
double gluon-gluon distribution. The remnant terms of sum in 
eq.~(\ref{solQCD}) are relatively small and can only increase the effect under 
consideration because they are positive.

The results of numerical calculations are presented on fig.~1 for the ratio:
\begin{equation}
\label{ratio}
R(x,t)~=~\Big(D_{p(QCD)}^{gg}(x_1,x_2,t) \Big/ D_p^{g}(x_1,t)D_p^{g}(x_2,t)(1 - x_1
- x_2)^2\Big)\Big|_{x_1=x_2=x}.
\end{equation}
Here one should note that the momentum conserving phase space factor $(1 - x_1 - x_2)^2$
is introduced in eq.~(\ref{ratio}) instead of $(1 - x_1 - x_2)$ usually used. 
The reason is simple: this factor was introduced in eq.~(\ref{factoriz}), generally
speaking, ``by hand'' in order to ``save'' the momentum conservation law, i.e. in
order to make the product of two single distributions is equal to zero smoothly at
$x_1 + x_2 = 1.$ However the generalized QCD evolution equations demand  higher
power of $(1 - x_1 - x_2)$ at $ x_1 + x_2 \rightarrow 1$: only the phase space
integrals in eqs. (\ref{solution1}) and (\ref{solQCD}) give
$$
\int\limits_{x_1}^{1-x_2} dz_1
\int\limits_{x_2}^{1-z_1}dz_2~ =~(1 - x_1 - x_2)^2/2 .$$
In fact this power must depend on $t$ increasing with its growth  as this takes
place for single distributions   at $ x \rightarrow 1$~\cite{dok, dok2}. The
asymptotic behaviour of two-particle fragmentation functions at 
$ x_1 + x_2 \rightarrow 1$ was investigated, for instance, 
in ref.~\cite{vendramin} with the
similar result. Our numerical calculations support this assertion also:
the power of $(1 - x_1 - x_2)$ for the perturbative QCD gluon-gluon correlations
is  higher than 2 and increases with $t(Q)$ as one can see from fig.~1 However
the introduced factor $(1 - x_1 - x_2)^2$ has not an influence practically
on the ratio under consideration  in the region of small $x_1, x_2$. And namely
this region, in which multiple interactions can contribute to the cross section
visibly, is interesting from experimental point of view. Fig.~1 shows that
at the scale of CDF hard process ($\sim 5$ GeV)
the ratio (\ref{ratio}) 
 is nearly 10$\%$  and increases right up to 30$\%$ at the LHC 
scale ($\sim 100$ GeV) for the longitudinal momentum fractions $x \leq 0.1$
accessible to these measurements. For the finite longitudinal momentum
fractions $x \sim 0.2 \div 0.4$ the correlations are large right up to 90$\%$ .
They become important in more and more $x$ region   with the growth of $t$
in accordance with the predicted QCD asymptotic behaviour.

The correlation effect is strengthened insignificantly (up to 2$\%$)
for the longitudinal momentum fractions $x \leq 0.1$ when
starting from the slightly lower value $Q_0 = 1$ GeV (early used by CTEQ 
Collaboration). We conclude also that $R(x,t)\rightarrow const$ at $x \rightarrow 0$
most likely, calculating this ratio ($\simeq 0.1$) at $x_{min} = 10^{-4}$.

Seemingly the correction to the double gluon-gluon distributions at the CDF
scale can be smoothly absorbed by uncertainties in the $\sigma_{eff}$ increasing
the transverse effective size $r_p$ by a such way. But this augmentation is still 
not enough to solve a problem of the relatively small value of $r_p$ with 
respect to the proton radius without  
nontrivial correlation effects in transverse space~\cite{trelani,trelani2}.

In summary, the numerical estimations show that the leading logarithm perturbative
QCD correlations are quite comparable with the factorized distributions. With
increasing a number of observable multiple collisions (statistic) the more precise
calculations of their cross section (beyond the factorization hypothesis) will be
needed also. In order to obtain the more delicate their characteristics
(distributions over various kinematic variables) it is desirable to implement the
QCD evolution of two-parton distribution functions in some Monte Carlo event
generator as this was done for single distributions within, for instance,
PYTHIA~\cite{pythia}.

\noindent
{\it Acknowledgements} 

\noindent
It is pleasure to thank V.A.~Ilyin drawing the attention to
the problem of  double parton scattering.
Discussions with E.E.~Boos, M.N.~Dubinin, I.P.~Lokhtin, S.V.~Molodtsov, 
L.I.~Sarycheva, V.I.~Savrin, T.~Sjostrand, D.~Treleani 
and G.M.~Zinovjev are gratefully acknowledged. Authors are specially thankful to 
A.S.~Proskuryakov for the valuable help in adapting CTEQ Evolution package.
This work is partly supported 
by grant N 04-02-16333 of Russian Foundation for Basic Research.

\newpage
\begin{figure}[hbtp] 
\begin{center} 
\makebox{\epsfig{file=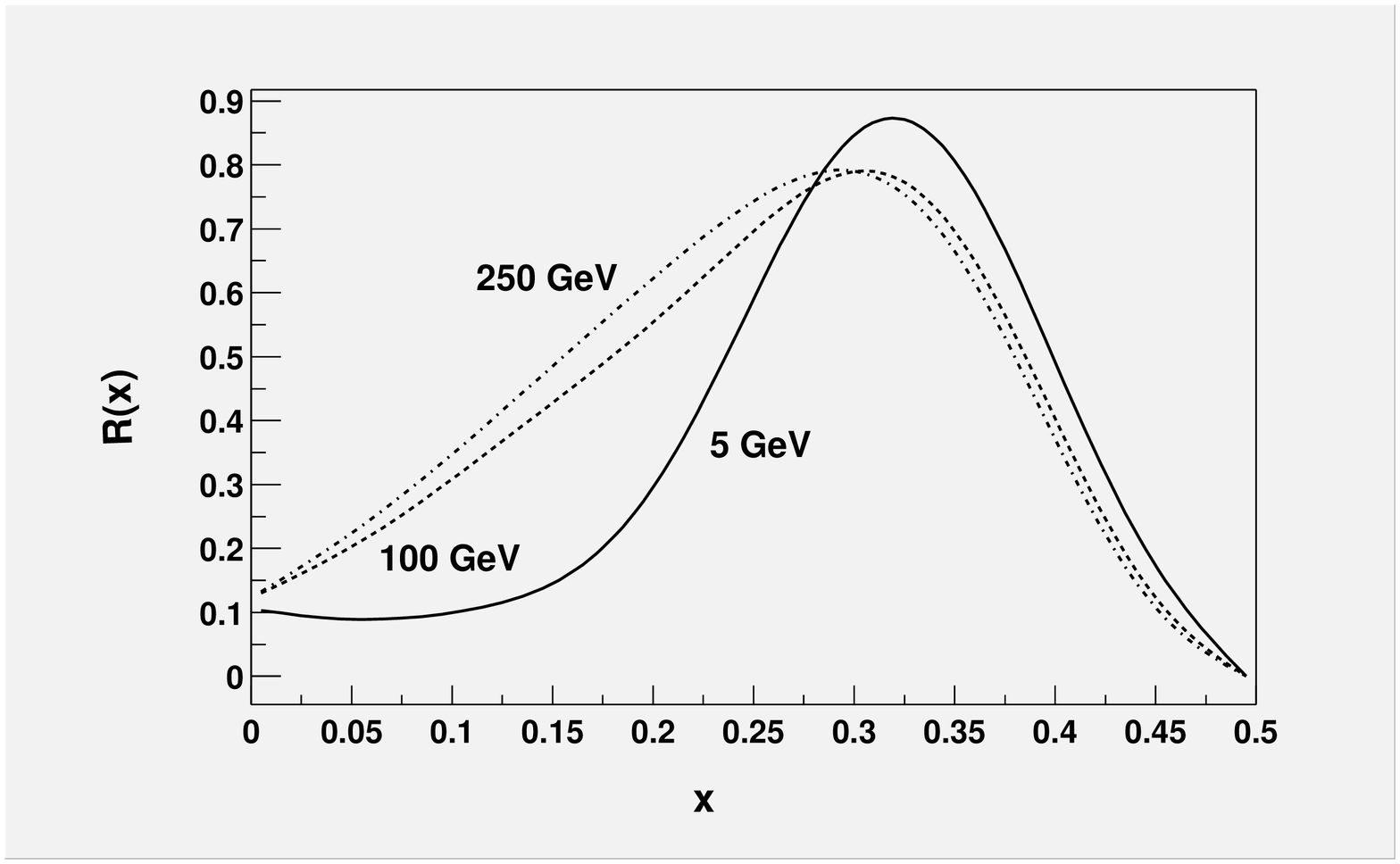, height=170mm, angle=270}}   
\vskip 1cm 
\caption{The ratio of perturbative QCD correlations to the factorized component for
the double gluon-gluon distribution in the proton as a function of $x = x_1 = x_2$
for three values of $Q$ = 5(solid), 100(dashed), 250(dash-dotted) GeV.}  
\end{center}
\end{figure}


\newpage 

\begin{thebibliography}{99} 
\bibitem{cdf} CDF Collab., F. Abe et al., Phys. Rev. D 56 (1997) 3811.
\bibitem{landshoff} P.V. Landshoff and J.C. Polkinghorne, Phys. Rev. D 18 (1978)
3344.
\bibitem{goebel} C. Goebel, F. Halzen and D.M. Scott, Phys. Rev. D 22 (1980) 2789.
\bibitem{takagi} F. Takagi, Phys. Rev. Lett. 43 (1979) 1296.
\bibitem{paver} N. Paver and D. Treleani, Nuovo Cimento A 70 (1982) 215.
\bibitem{humpert} B. Humpert,  Phys. Lett. 131 B (1983) 461.
\bibitem{odorico} B. Humpert and R. Odorico, Phys. Lett. 154 B (1985) 211.
\bibitem{sjostrand} T. Sjostrand and M. van Zijl, Phys. Rev. D 36 (1987) 2019.
\bibitem{trelani} G. Calucci and D. Treleani,
Nucl. Phys. B (Proc. Suppl.) 71 (1999) 392.
\bibitem{trelani2} G. Calucci and D. Treleani, Phys. Rev. D 60
(1999) 054023.
\bibitem{del}A. Del Fabbro and D. Treleani, Phys. Rev. D 61 (2000) 077502;\\
 Nucl. Phys. B (Proc. Suppl.) 92 (2001) 130. 
\bibitem{sjostrand2} T. Sjostrand and P.Z. Scands , hep-ph/0402078.
\bibitem{snig03} A.M. Snigirev, Phys. Rev. D 68 (2003) 114012.
\bibitem{gribov} V.N. Gribov and L.N. Lipatov, Sov. J. Nucl. Phys. 15 (1972) 438; 
15 (1972) 675.
\bibitem{lipatov} L.N. Lipatov, Sov. J. Nucl. Phys. 20 (1974) 94.
\bibitem{dok}Yu.L. Dokshitzer, Sov. Phys. JETP 46 (1977) 641.
\bibitem{kirschner} R. Kirschner, Phys. Lett. 84 B (1979) 266.
\bibitem{snig} V.P. Shelest, A.M. Snigirev and G.M. Zinovjev, Phys. Lett. 113 B
(1982) 325;\\
 Sov. Theor. Math. Phys. 51 (1982) 523.
\bibitem{altarelli} G. Altarelli and G. Parisi, Nucl. Phys. B 126 (1977) 298.
\bibitem{konishi} K. Konishi, A. Ukawa and G. Veneziano, Phys. Lett. 78 B (1978)
243;\\
Nucl. Phys. B 157 (1979) 45. 
\bibitem{collins} J.C. Collins, D.E. Soper and G. Sterman, in: Perturbative
QCD, edited by A. Mueller (World Scientific, Singapore, 1989). 
\bibitem{cteq} CTEQ Collab., H.L. Lai et al.,
Eur. Phys. J. C 12 (2000) 375. 
\bibitem{snig2} V.P. Shelest, A.M. Snigirev and G.M. Zinovjev, JINR Communication,
E2-82-194, Dubna (1982); Preprint ITP-83-46E, Kiev (1983), unpublished.
\bibitem{puhala} M.J. Puhala, Phys. Rev. D 22 (1980) 1087.
\bibitem{pumplin} J. Pumplin et. al., JHEP (2002) 0207:012.
\bibitem{Evolve} http://www.phys.psu.edu/~cteq/
\bibitem{dok2} Yu.L. Dokshitzer, D.I. Dyakonov and S.I. Troyan, Phys. Rep. 58 (1980)
269.
\bibitem{vendramin} I. Vendramin, Nuovo Cimento 62 A (1981) 21.
\bibitem{pythia} T. Sjostrand, Comp. Phys. Com. 135 (2001) 238.
\end{thebibliography}
\end{document}